\documentclass[
reprint,
superscriptaddress,
 amsmath,amssymb,
 aps, prx,
floatfix,
]{revtex4-2}
\usepackage{siunitx}[per-symbol] 
\usepackage{graphicx}
\usepackage{dcolumn}
\usepackage{bm}

\newcommand{\nitride}{Si\textsubscript{3}N\textsubscript{4}}

\newcommand{\oxide}{SiO\textsubscript{2}}

\newcommand{\crosstalk}{\SI[uncertainty-mode=compact-marker]{-50.8(1.3)}{\dB}}
\newcommand{\pcrosstalk}{\SI[uncertainty-mode=compact-marker]{50.8(1.3)}{\dB}}
\newcommand{\fibercrosstalk}{\SI[uncertainty-mode=compact-marker]{-60.6(2.5)}{\dB}}

\newcommand{\repumper}{\ensuremath{\text{D}_{3/2} \!\leftrightarrow\! \text{P}_{1/2}}}

\newcommand{\magnification}[1]{\ensuremath{{#1}\kern-.1em\times}}

\newcommand{\afrl}{\affiliation{Air Force Research Laboratory, Rome, NY}}
\newcommand{\tchngs}{\affiliation{Technergetics, Utica, NY}}
\newcommand{\booz}{\affiliation{Booz Allen Hamilton, Rome, NY}}
\newcommand{\griffiss}{\affiliation{Griffiss Institute, Rome, NY}}
\newcommand{\murray}{\affiliation{Murray Associates of Utica, Utica, NY}}
\newcommand{\rit}{\affiliation{Rochester Institute of Technology, Rochester, NY}}
\newcommand{\rfsuny}{\affiliation{Research Foundation for the State University of New York, Albany, NY}}
\newcommand{\NRL}{\affiliation{U.S. Naval Research Laboratory, Washington DC}}
\newcommand{\AIM}{\affiliation{AIM Photonics, Albany, NY}}
\newcommand{\HRL}{\affiliation{HRL Laboratories, Malibu, CA}}
\begin{document}

\title{Low-Crosstalk, Silicon-Fabricated Optical Waveguides for Laser Delivery to Matter Qubits}

\author{Clayton L. Craft}
\email{clayton.craft.1@afrl.af.mil}
\afrl

\author{Nicholas J. Barton}
\murray

\author{Andrew C. Klug}
\tchngs

\author{Kenneth Scalzi}
\tchngs

\author{Ian Wildemann}
\griffiss

\author{Pramod Asagodu}
\griffiss

\author{Joseph D. Broz}
\HRL

\author{Nikola L. Porto}
\griffiss

\author{Michael Macalik}
\booz

\author{Anthony Rizzo}
\afrl

\author{Garrett Percevault}
\afrl

\author{Christopher C. Tison}
\afrl

\author{A. Matthew Smith}
\afrl

\author{Michael L. Fanto}
\afrl

\author{James Schneeloch}
\afrl

\author{Erin Sheridan}
\afrl

\author{Dylan Heberle}
\tchngs

\author{Andrew Brownell}
\murray

\author{Vijay S.S. Sundaram}
\rit

\author{Venkatesh Deenadayalan}
\rit

\author{Matthew van Niekerk}
\rit

\author{Evan Manfreda-Schulz}
\rit

\author{Gregory A. Howland}
\rit

\author{Stefan F. Preble}
\rit

\author{Daniel Coleman}
\rfsuny
\AIM

\author{Gerald Leake}
\rfsuny
\AIM

\author{Alin Antohe}
\rfsuny
\AIM

\author{Tuan Vo}
\rfsuny
\AIM

\author{Nicholas M. Fahrenkopf}
\rfsuny
\AIM

\author{Todd H. Stievater}
\NRL

\author{Kathy-Anne Brickman-Soderberg}
\afrl

\author{Zachary S. Smith}
\afrl

\author{David Hucul}
\afrl

\date{\today}

\begin{abstract}
Reliable control of quantum information in matter-based qubits requires precisely applied external fields, and unaccounted for spatial cross-talk of these fields between adjacent qubits leads to loss of fidelity.
We report a CMOS foundry-produced, micro-fabricated silicon nitride (\nitride{}) optical waveguide for addressing a chain of eight, unequally-spaced trapped barium ions with crosstalk compatible with scalable quantum information processing. The crosstalk mitigation techniques incorporated into the chip design result in a reduction of the measured optical field by at least \pcrosstalk{} between adjacent waveguide outputs near \SI{650}{nm} and similar behavior for devices designed for \SIlist{493;585}{\nm}. The waveguide outputs near \SI{650}{\nm}, along with a global laser near \SI{493}{\nm} were used to laser-cool a chain of eight barium-138 ions, and a camera imaged the resulting fluorescence at \SI{493}{\nm}.
\end{abstract}

\maketitle


\section{\label{sec:intro}Introduction}
Matter-based quantum bits (qubits) controlled with lasers are promising hosts of quantum information because these quantum memories (\emph{e.g.}\ trapped ions \cite{IonQManyQubits, QuantinuumRacetrack, RackMountQC}, neutral atoms \cite{evered2023, bernien2017probing, barnes2022}, nitrogen vacancy centers \cite{NVcenterGoldman, bradley2022robust}, and quantum dots \cite{millsPetta2022}) can possess long coherence times \cite{WangP:2021, young2020half, abobeih2022fault}, have high-fidelity single qubit control and two qubit gates \cite{Clark:2021, ballance2016high, gaebler:2016, an:2022} and can be networked to increase their processing power \cite{monroe:2014, stephenson:2020, covey2023quantum, humphreys2018deterministic}. 
Well-focused laser delivery to many closely spaced qubits with visible wavelength transitions, especially when irregularly spaced, presents an optical engineering challenge \cite{kwon:2024, bluvstein:2024}.
The use of bulk optical components to adjust the angle and position of the laser beams are not readily scalable to larger systems.
Integrated, diffractive optics can deliver lasers to address qubits within widely-spaced trapping zones \cite{niffenegger:2020, mehta2017precise, vasquez:2023, hogle:2023}, and laser-written waveguides (in glass) have delivered light to multiple qubits in a single trapping zone with low crosstalk \cite{binaimotlagh2023guided, timpu2022laser}.
Foundry-compatible, Si-based waveguides with high index contrast have very low cross-talk between waveguide channels while maintaining efficient mode matching to closely spaced qubits for light delivery and collection.

\begin{figure*}
\includegraphics[width={7 in}]{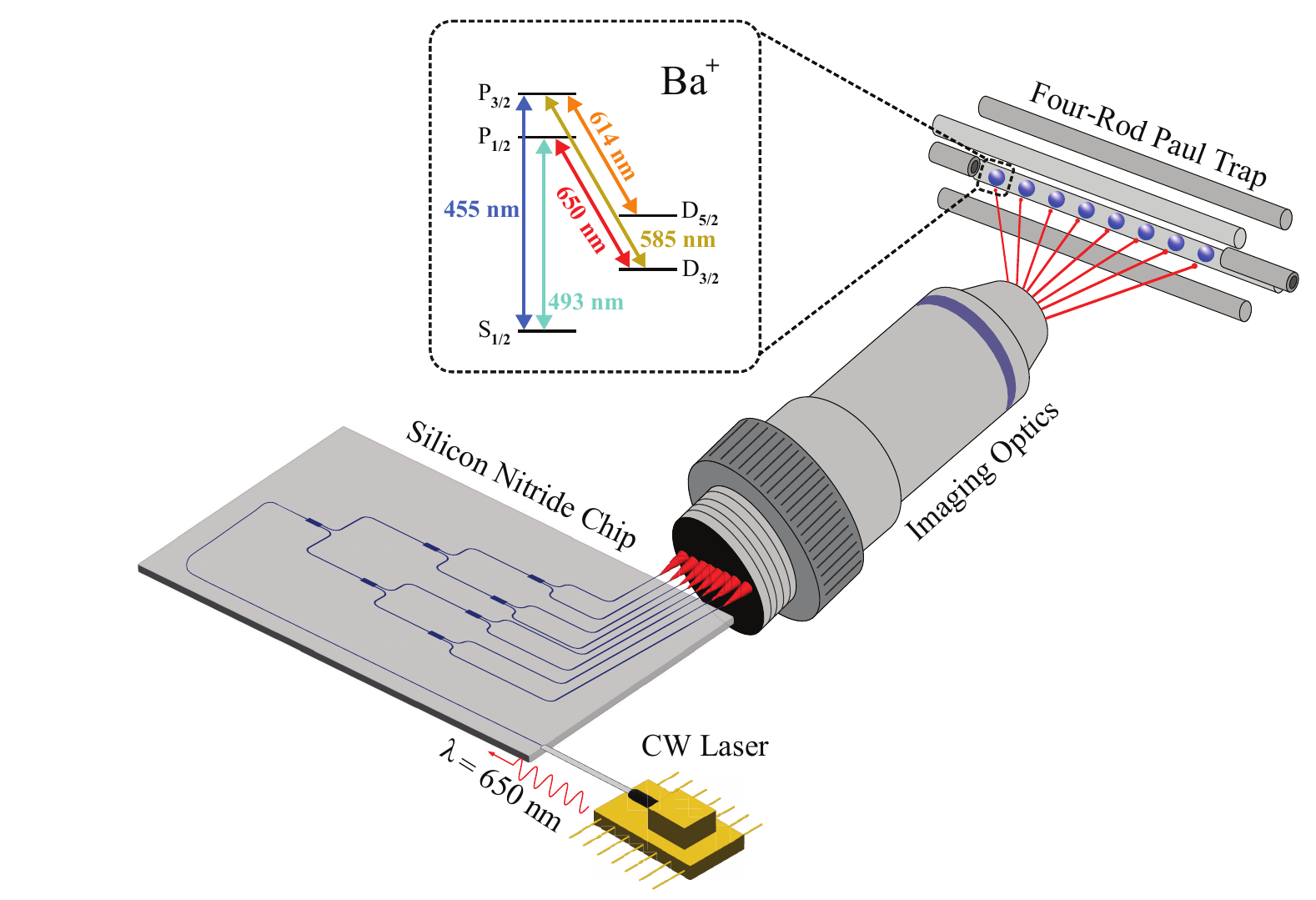}
\caption{Illustration of a photonic chip to deliver light to and collect light from an array of qubits. 
The waveguide edge couplers and lens system are selected to mode-match to individual qubits in the array.
This enables the low-crosstalk output light from the chip to be focused onto individual qubits as well as the efficient, parallel collection of each of their optical emissions. The use of low loss bends, trenches, and spot size conversion in a foundry-produced, high index contrast photonics chip results in low crosstalk between waveguide outputs.
} \label{fig:PrimaryFigure}
\end{figure*}

In this work, we detail our efforts to fabricate a silicon nitride-based (\nitride-based) waveguide that splits an input laser into multiple outputs for addressing an array of matter-based qubits while minimizing the crosstalk between the output channels. 
We verify this by directly measuring the crosstalk of the waveguide outputs near \SIlist{493;585;650}{\nm}.
As a demonstration of the viability of the waveguides, the separate outputs were used to perform Doppler cooling of a chain of eight $^{138}\text{Ba}^+$ ions by driving the \repumper{} repumping transition (see Fig. \ref{fig:PrimaryFigure}).
Addressing the ions in this way serves as a model application of this photonic chip technology.
This technology is extendable to individual qubit manipulations \cite{binaimotlagh2023guided, mehta2017precise},  mid-circuit measurements \cite{zhu2023interactive, foss2023experimental}, and quantum networking with efficient light collection \cite{carter:2024}.

\section{\label{sec:fab}Photonic Chip Design and Fabrication}

The requirement of interacting with closely spaced qubits drives the optimal design of our waveguide device.
In particular, we focused on ensuring minimal quantities of stray light leak through the photonic chip between the waveguide outputs and providing good mode matching to close but unequally spaced qubits in a single trapping zone.

\begin{figure}
\includegraphics[width = \columnwidth]{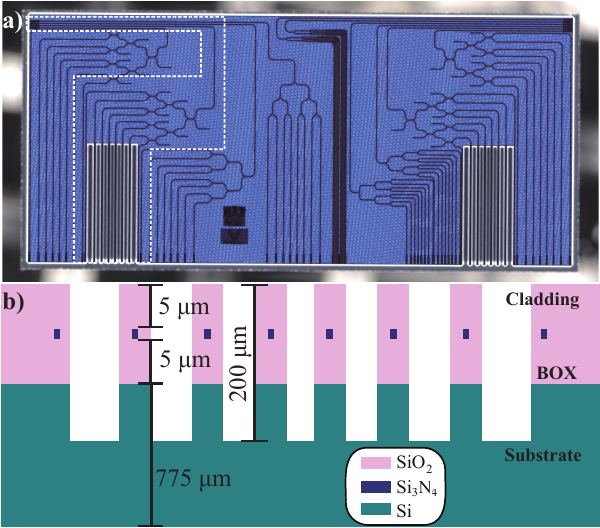}
\caption{\label{fig:widephotons}Photonic chip platform design and simulation.
\textbf{(a)} Microscope image of a photonic chip including several device variants.
The device outlined in dotted white (input on the top left) has multiple salient passive features for crosstalk mitigation, including trenches (bottom), output spot  size  conversion, and waveguide output spacing.
\textbf{(b)} The cross sectional area of the waveguide output (not drawn to scale for clarity).
The waveguide material (\nitride{}, dark blue) tightly confines the light.
The surrounding cladding (\oxide{}, pink) surrounds the top and sides of the waveguide and is separated from the silicon substrate (Si, green) by a buried oxide layer (BOX, pink).
The trenches extend fully through the cladding, BOX, and partially into the substrate, ensuring no cladding is left to support a weakly-guided slab mode.}
\end{figure}

The reduction in stray light primarily stems from strategically placed \ang{90} bends in the waveguides and trenches etched between the outputs as shown in Fig. \ref{fig:widephotons}.
The mode-matching capabilities come from tapering the waveguide outputs and patterning their spacing to reflect the relative spacing between the qubits.
There is a tradeoff between optimizing mode-matching capabilities and limiting crosstalk as increasing waveguide spacing reduces crosstalk but requires higher magnification imaging optics to match the ion pitch.
In addition, focusing the waveguide outputs through high magnification optics complicates matching the output mode field diameter to the ions.

\subsection{\label{sec:fab:foundry}Foundry Platform}

We fabricated the integrated photonic chips on \SI{300}{\mm} silicon-on-insulator (SOI) wafers using a commercial Complementary Metal-Oxide-Semiconductor (CMOS) foundry, AIM Photonics \cite{fahrenkopf:2019}.
Due to the band gap of silicon (\SI{1.1}{\electronvolt}), standard SOI waveguides absorb wavelengths below $\lambda\!\sim\!\SI{1100}{\nm}$, precluding their use for applications in the visible range. However, \nitride{}  waveguides deposited on silicon wafers provide a method to extend the spectral range to the near-UV region while maintaining compatibility with standard CMOS tooling \cite{sorace:2019, sacher:2023, smith:2023, sundaram:2022}. Our platform consists of \SI{150}{\nm} thick \nitride{} waveguides deposited on silicon wafers through low-pressure chemical vapor deposition (LPCVD). The chosen waveguide thickness realizes high modal confinement, enabling compact device footprints with small, low-loss bends \cite{sundaram:2022}.
A \SI{5}{\um} layer of buried oxide (BOX) isolates the waveguide mode from the silicon substrate, and \SI{5}{\um} of \oxide{} is deposited as the top cladding of the waveguide as shown in the cross-section schematic in Fig. \ref{fig:widephotons}(b)). The \nitride{} waveguides are patterned using \SI{193}{\nm} deep-UV argon-fluoride (DUV ArF) excimer-laser immersion lithography and then etched \cite{tyndall:2021,Tyndall:23}.

\subsection{\label{sec:fab:design}Waveguide Design}

\begin{figure}
    \includegraphics[]{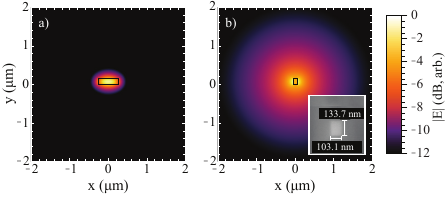}
    \caption{\label{fig:spotsize}
    \textbf{(a)} The simulated waveguide mode for the designed on-chip routing. The black box indicates the spatial extent of the \SI{500}{\nm} by \SI{150}{\nm} waveguide in the simulation ($\lambda$ = 650 \SI{650}{\nm}, TE mode).
    \textbf{(b)} The output edge coupler after the spot size converter to taper the waveguide in a) to a \SI{100}{\nm} by \SI{150}{\nm} waveguide, enlarging the output mode field diameter. Inset is a scanning electron microscope image of the output facet of the fabricated chip.}
\end{figure}

Finite difference eigenmode (FDE) simulations indicated that the optimal waveguide width for \SI{650}{\nm} was \SI{500}{\nm}, which ensures that the design is below the single mode cutoff while maintaining a high confinement and low modal sidewall overlap.
Waveguide dimensions for other wavelengths were similarly developed by simulations to balance the single mode cutoff and sidewall overlap. 
The simulated electric field profile of the fundamental transverse electric (TE) mode for a nominal waveguide is shown in Fig. \ref{fig:spotsize}(a).

Due to the high modal confinement of the waveguide geometry, compact \SI{10}{\um} radius bends with simulated losses below \SI{0.1}{\decibel}/\ang{90} bend are used to route the photonic circuits.
The fundamental TE mode has a measured loss of \SI[per-mode=symbol]{1.7}{\decibel\per\centi\meter} though a long section of waveguide. 
Light was equally distributed to the outputs using a splitter tree. Directional coupler based 4-port splitters\cite{Lu:2013, Lalau-Keraly:2013} were tested along with multi-mode interference (MMI) y-junctions for a differences in crosstalk mitigation, especially in regards to polarization stability as the TM mode was not confined in the waveguide as well as the TE mode. However, they both display a large polarization dependency in the transmission efficiency as well as the ratios between output waveguides. The designs were simulated and optimized using Ansys Lumerical.

The fiber-chip and chip-free space interfaces use inverse tapered spot size converters (SSCs) to expand the highly confined waveguide mode to expand the highly confined waveguide mode so that focusing the mode through an objective produces diffraction limited spots at the ion positions\cite{Almeida:03}.
The \SI{650}{\nm} waveguide width is linearly tapered from \SI{500}{\nm} to \SI{125}{\nm} over a length of \SI{100}{\um} such that the mode is adiabatically expanded without abrupt modal mismatch interfaces.
The resulting output mode simulated is shown in Fig. \ref{fig:spotsize}(b) with an inset scanning electron microscope (SEM) image of the \nitride{} SSC tip at the chip facet.

\subsection{\label{sec:fab:architecture}Photonic Circuit Architecture}
Careful consideration of the chip architecture enables design choices that suppress unwanted output light, exemplified by the device characterized in this work, outlined in white dashes in Fig. \ref{fig:widephotons}(a).
In addition to the intended in-waveguide propagation, weakly-guided slab modes support transmission of undesired light through the device in the cladding.
Of particular concern are areas where light that does not couple into the waveguides is strongly concentrated, such as the input facet of the photonic chip, and at the splitter interfaces.
Two sets of turns ensure this slab mode does not contaminate the output light.
First, a pair of \ang{90} radial bends ensures stray input light does not couple into the splitter tree structure.
Second, a \ang{90} radial bend directs the output perpendicular to scattered light at both the input and splitter junction(s).
To further extinguish slab modes, we deep-etched air trenches (extending into the substrate) between output waveguides to ensure light in this region is not guided. Instead the light diverges rapidly from the beginning of the trench for minimal coupling to the qubits.

This architecture allows control over the spacing of the waveguide output channels and, via SSCs, the spot size at the output facet.
Combined with the effective refractive index of the waveguide, the spot size also sets the output NA of the chip.
Meanwhile, the qubit array will have its own spacing, desired spot size, and target NA (for light collection).
These parameters are not fully independent; the chip and qubit spacings are related by the imaging system magnification, as are the corresponding spot sizes.
In contrast, the NA is inversely proportional to the spot size
\footnote{The NAs at the chip and qubit are also related by the magnification, but this constraint is redundant and can be obtained by combining the earlier constraints of spot size, NA, and magnification.}.
Altogether, these form 7 parameters (spot size, spacing, and NA at both the chip and the qubit, and the imaging system magnification) and 4 constraining relationships, leaving 3 freely chosen parameters available during system design.
The qubit platform likely sets the qubit spacing,
but the other two may be chosen to enable experimental goals.
For example, increasing the ratio of the spacing to the spot size allows for greater mode falloff with distance, potentially improving crosstalk, and choosing a large demagnification factor would scale down any errors in positioning the chip relative to the qubits.
Alternatively, choosing a pair of NAs to enable efficient light collection without requiring an unwieldly imaging system may have a higher experimental priority.

For the device pictured in Fig. \ref{fig:widephotons}(a), the spacing between outputs correlated to the spacing between ions in a RF Paul Trap \cite{meyrath:1998} and scaled to 5-10 times larger than the spacing between ions in the trap.
This allows for \magnification{<1} magnification optics to reduce the size of the beam profile for more efficient light delivery.

The high modal confinement of the waveguide geometry afforded compact, \SI{10}{\micro\meter} radius bends to route the photonic circuits with simulated losses below \SI{0.1}{\dB}/\ang{90} bend.
This allows for turning the waveguides such that the output SSCs face an adjacent facet of the chip rather than the opposite facet.  This geometry ensures that stray weakly-guided light in slab modes excited by the fiber-chip input interface does not exit from the same side of the chip as the output waveguides.
Furthermore, deep-etched air trenches extending from the top surface of the chip, down past the waveguides into the handle wafer and extending to the edge facet are fabricated between each waveguide output. This helps to ensure that light cannot be guided between waveguides.
The spacing between outputs correlates to the relative spacing between ions in our RF Paul Trap\cite{meyrath:1998} but is scaled to \magnification{\sim5-10} larger than the spacing between ions.
The waveguide spacing was chosen to match the magnification (\magnification{<1}) and NA of the microscope objective in our experiment.

\section{\label{sec:bench}Direct Optical Characterization}
To best characterize the limiting performance of the fabricated photonic chip, we directly measured its output intensity using a microscope objective and a scanning slit in the objective's image plane. Our measurement technique includes residual crosstalk from the waveguide outputs and any residual stray light, such as slab modes, guided through the photonics chip.
Incorporating this stray light into our crosstalk metric worsens the apparent performance from a true crosstalk-only measurement, and encompasses stray light that can lead to loss of qubit fidelity.
This more conservative crosstalk metric will be used throughout the paper.

\begin{figure}
 \includegraphics[width=3.4in]{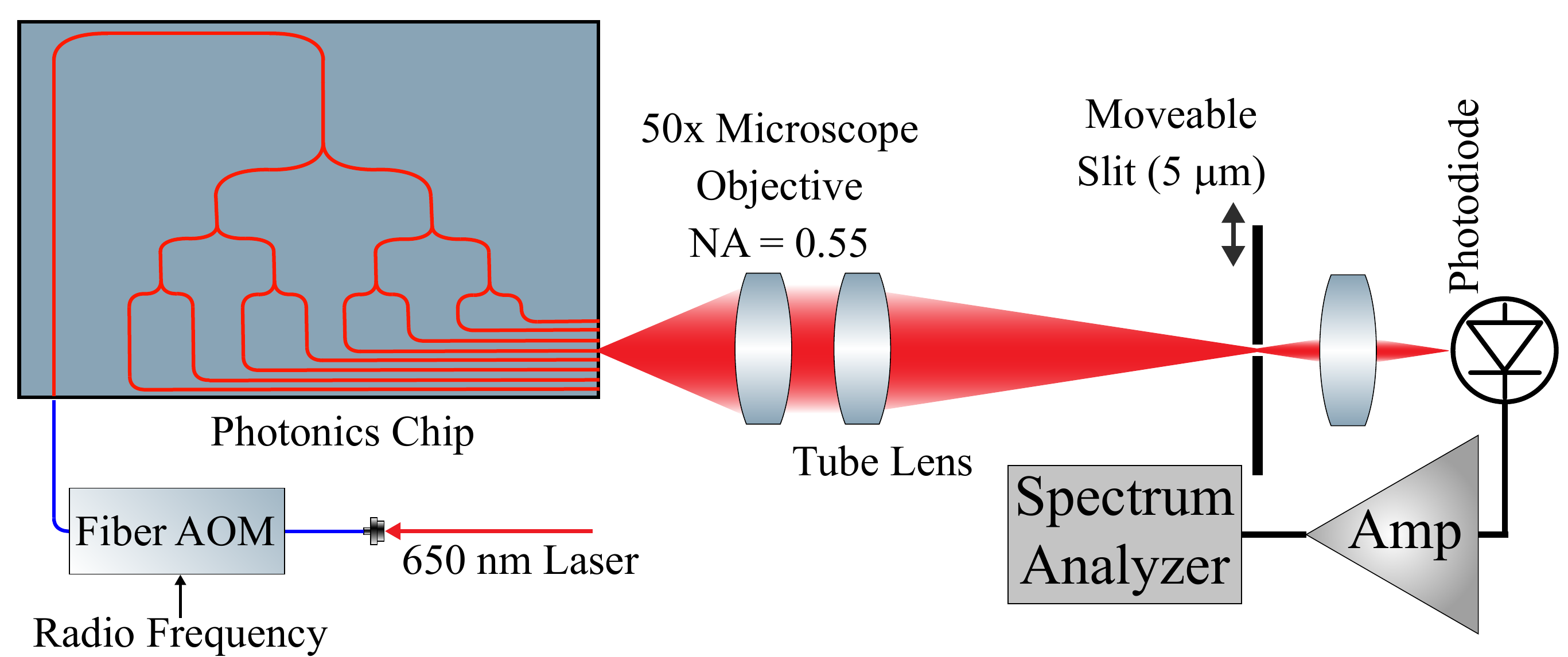}
\caption{Optical setup for characterizing waveguide output and crosstalk.
A fiber-coupled acousto-optic modulator (AOM) amplitude modulates a laser (tuned to \SI{650}{\nm} near the \repumper{} transition in Ba$^{+}$ ions) at \SI{28.1}{\kHz}. 
The resultant fiber output couples directly into the photonic chip, which divides the light into eight outputs whose spacing matches a chain of eight trapped barium ions (up to a scale factor).
The output of the chip is collimated using a NA = 0.55, \magnification{50} objective and is imaged using a \SI{200}{\mm} tube lens.
The light passes through a movable \SI{5}{\um} width slit attached to a high-gain photodiode.
A spectrum analyzer extracts the signal component at the modulation frequency.
} \label{fig:setup}
\end{figure}

\subsection{Characterization Technique}\label{subsec:directdesign}
Measurements utilized the configuration shown in Fig. \ref{fig:setup}.
A fiber acousto-optic modulator (AOM) modulates the amplitude of the input laser so that the signal light can be isolated from any background.
A 0.55 NA objective with a \SI{200}{\mm} tube lens formed a \magnification{50}-magnified image.
A \SI{5}{\um} width, \SI{3}{\mm} height slit moving along a linear positioning stage at \SI{1}{\um} steps transmits a small slice of the image.
The intensity of this section of light is detected by a variable-gain photodiode that travels in conjunction with the slit and from which the amplitude-modulated component is isolated using a signal analyzer.
This method convolves the magnified image with the slit, so the original waveguide output profile is reconstructed using Mathematica’s built-in deconvolution algorithm\cite{Mathematica} using a top-hat function as the kernel. Because slit size is much smaller than the image of the waveguide outputs and the kernel of a slit is simple, these deconvolutions accurately reflect the transversely-integrated intensity profile.

Since the range of motion of the slit is limited to \SI{10}{\mm}, multiple scans were combined.
Due to gradual fiber drift over several hours, some decoupling is expected across the full range of slit motion.
Taking shorter ($\approx$\SI{30}{\minute}) scans allows frequent re-optimization of input couplings,
as evidenced by a maximum \SI{1}{\decibel} peak height difference between temporally adjacent scans.
To compensate for any residual output intensity drift, overlapping peaks were normalized to each other during recombination.
The above measurement can be limited by misalignment, scattering, and reflections from the optical imaging system.

A two-dimensional distribution of light was obtained by scanning a single-mode optical fiber across the output face to directly couple light into the fiber, avoiding the use of optics. 
Deconvolving the measured intensity distribution with the single-mode fiber kernel is complicated by unknown phase and edge effect contributions to the fiber's transfer function.
Since the mode field diameter of the fiber is non-negligible in size in comparison to the chip output, this perturbation is expected to be significant, but is presented without deconvolution to avoid introducing large artifacts into the reported profile.
In addition to providing a two-dimensional intensity profile, the larger kernel increases the measurement dynamic range by collecting a larger portion of the output beam.

\subsection{\label{sec:bench:result}Characterization Results}

\begin{figure}
 \includegraphics[width=3.4in]{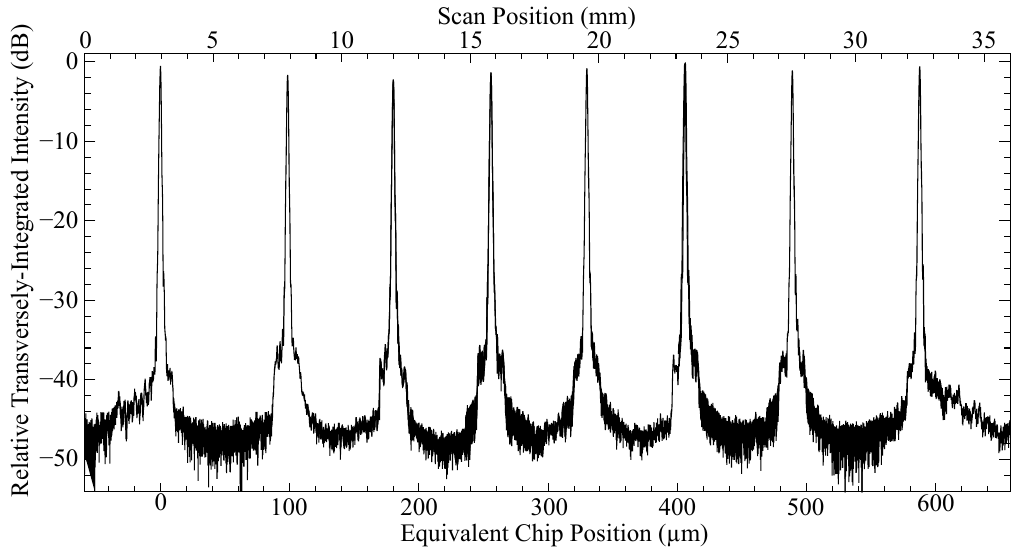}
\caption{Measured photonic chip output profile using the \magnification{50} imaging system and apparatus shown in Fig. \ref{fig:setup}.
The slit assembly samples the image plane at the indicated position (top axis).
This corresponds to the equivalent chip position (bottom axis) inferred from comparing the output peak locations to the chip design specifications.
Displayed is a composite profile across the chip, show the relative transversely-integrated intensity profile after deconvolution (see text for details).
The markedly sharper falloff between waveguide channels is due to the trenches, which are not present on the outside edges.
} \label{fig:datastitch}
\end{figure}

The full light distribution of the photonic chip output as measured by the scanning slit is given in Fig. \ref{fig:datastitch}.
Trenches were etched in the regions between two waveguides, leaving the outer two with no trenches towards the outside.
The impact of trenching is clearly evident as the intensity of these outer edges falls off more slowly than the sharply-decreasing intensity between the waveguides.
This assymmetry provides a demonstration of the impact trenching has on mitigating crosstalk.

\begin{figure}
  \centering
  \includegraphics[width=3.4in]{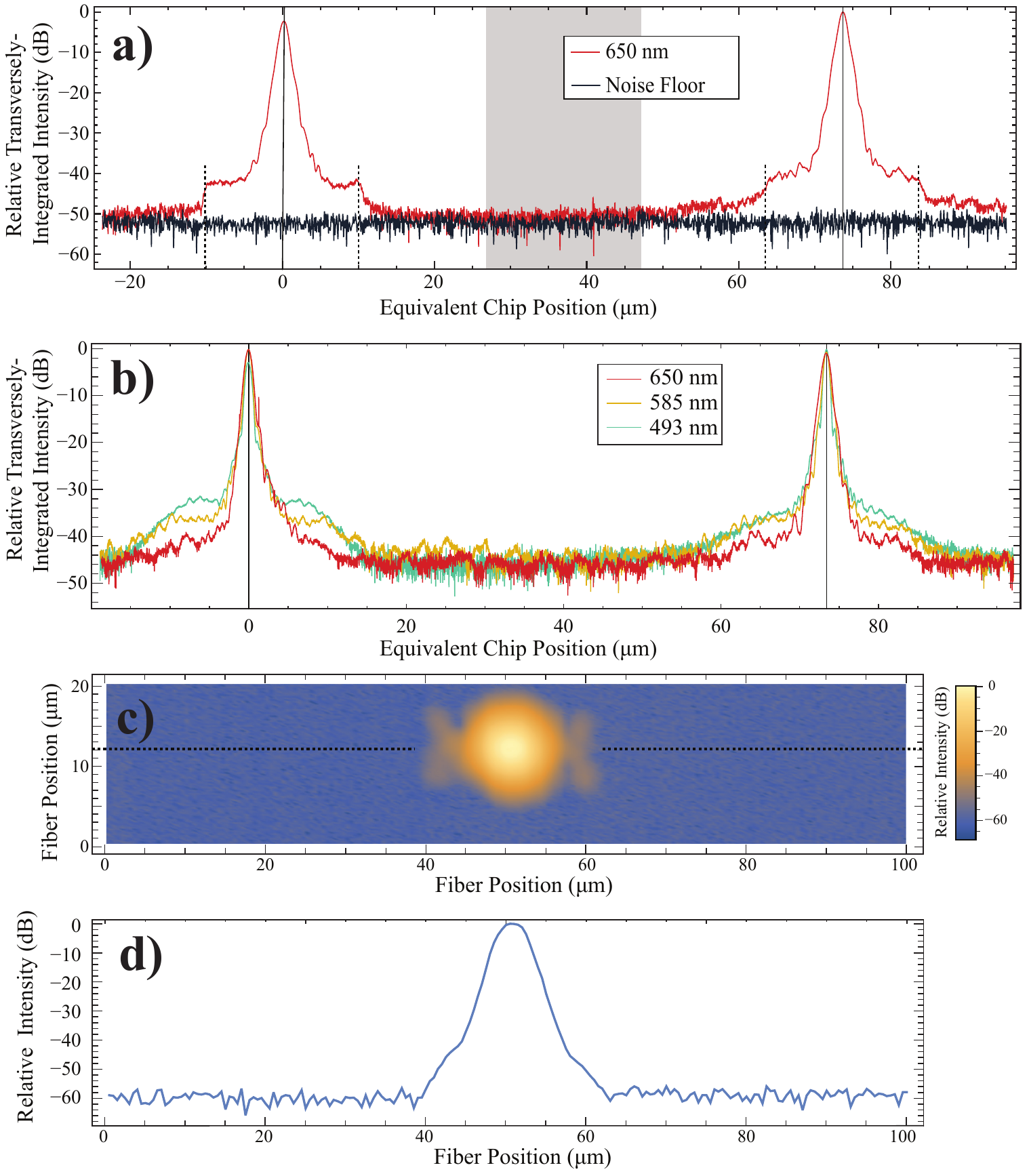}
\caption{Crosstalk measurements of waveguide outputs.
\textbf{(a)} The relative transversely-integrated intensity (red) across the two closest-spaced waveguides is compared to the noise floor detected when the laser is off.
The vertical lines refer to the peak centers (\SI{73.4}{\micro\meter} apart).
The measured crosstalk of \crosstalk{} compares the central peak height to the mean of the points in the shaded region.
The shaded region shows the extent of the cladding to scale,
demonstrating that trenching produces a sharp falloff between outputs.
\textbf{(b)} The relative transversely-integrated intensity distributions for three separate devices designed for transmitting \SIlist{493;585;650}{\nm} lasers (cyan, yellow, red) show similar crosstalk performance.
These devices do not have trenching and have a correspondingly less sharp falloff than their equivalents in (a),
and do not reach the lower crosstalk values of trenched devices.
\textbf{(c)} A direct-fiber scan provides a 2D profile of a trenched waveguide output, with a cut-though at the height of the waveguide maximum indicated by a dashed line (see Sec. \ref{subsec:directdesign}).
This isolates imaging aberration, scatter, and internal reflection effects from the intrinsic chip performance,
and improves the measurement dynamic range in our system.
\textbf{(d)} The relative intensity along the dashed line shown in c) highlights the measured crosstalk between the peak and the background, well away from the waveguide, as \fibercrosstalk{}.
} \label{fig:fiberscan}
\end{figure}

Since there is no active control to directly measure the unintended power distribution at the output, the area in between peaks provides a measurable stand-in,
and forms an upper bound because the waveguide mode is expected to fall off with increasing distance.
To find the most accurate measurement of crosstalk, excess scatter from light outside of the waveguide plane was mitigated by reducing the height of the slit.
The commercially available slit was modified by painting over the top and bottom which reduced the height to \SI{1.6}{\mm}.
This is much larger than the waveguide spot in the image plane, and the clipped light would not spatially overlap with a qubit in the envisioned use.
The ratio between the peak maximum and the average of 1000 points between two peaks shown in the shaded region of Fig. \ref{fig:fiberscan}(a) was \crosstalk{}.
This was extracted for the center and thus most tightly spaced waveguide pairs,
which are expected to have the largest crosstalk.

A comparison of the waveguide across \SIlist{493;585;650}{\nm} is shown in Fig. \ref{fig:fiberscan}(b).
Each trace shows similar crosstalk.
Although the final architecture uses trenches between the waveguide outputs, the only configuration that was available and consistent for all of these wavelengths does not include trenching.
We expect that the sharp cutoff trenching provides will not depend on wavelength.

The 2D direct-fiber scan in Fig. \ref{fig:fiberscan}(c) shows a ratio of maxmimum intensity to the background away from the waveguide of \fibercrosstalk{}.
It is likely that the discrepancy between the crosstalk values measured from the scanning slit method and direct coupling into a single mode fiber is due to coupling and edge effects inherent to the fiber-scan method as discussed above.

\section{\label{sec:ion}Demonstration: Laser-cooling trapped ions}
The development of this photonic circuit was motivated by separate addressing of any ion or combination of ions within a single trapping region.
A full, individually-addressed qubit control system would require active control of each of the splitter tree outputs.
However, the viability of the light delivery portion of such a system is demonstrated using this fully-passive device by providing repumping light to the ions as part of a Doppler cooling cycle.
The scattered cycling transition photons are detected by an Electron Multiplying Charge Coupled Device (EMCCD).

\subsection{Ion Demonstration Technique}
The output of commercial lasers used for the barium Doppler cooling transitions was used to illuminate the ions via two separate beampaths to provide cooling along all principal trap axes.
The laser wavelengths for \qtylist{493;650}{\nano\meter} (\qtylist{607.42625;461.31185}{\tera\hertz} respectively) were calibrated to to the R(56) 32-0 iodine transition near \SI{532}{\nm} (\SI{563.25965}{\tera\hertz}).
One laser path was aligned along the chain while the other was $\sim$\ang{45} to the chain of ions.
To account for the polarization dependency from the magnetic quantum numbers, an external magnetic field of a few gauss was applied to provide magnetic field components along the parallel and perpendicular directions to the laser's magnetic field polarization.

\begin{figure}
  \centering
  \includegraphics[width=3.4in]{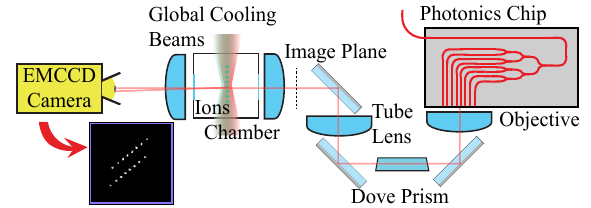}
\caption{Schematic of the trapped-ion and photonic chip setup.
The photonic chip produces a closely and unequally spaced outputs matching an eight-ion barium chain after passing through an imaging system with net \magnification{0.187} magnification.
A Dove prism rotates the resultant image to align with the ion chain.
(Inset) A fluorescence image at \SI{650}{\nm} of an eight-ion chain (lower grouping) is illuminated by \SIlist{493;650}{\nm} global cooling beams.
Simultaneously, the outputs of the decoupled photonic chip (upper grouping) are imaged on the camera at low output power.
This illustrates the agreement of the waveguide spacing as fabricated and imaged to the spacing of the chain in our four-rod Paul trap.
Aligning the chip output to the ions completes the laser cooling cycle to produce the fluorescence image in Fig. \ref{fig:PrimaryFigure} and described further in Sec \ref{sec:ion:result}.
} \label{fig:iondiagram}
\end{figure}

A four-rod RF Paul trap (\SI{34}{\kilo\hertz} axial secular frequency and \SI{152}{\kilo\hertz} lowest radial frequency) was used to confine crystallized barium ions generated via a sample of BaCl$_2$ directly ionized by laser ablation. Light scattered from the ions are collected by a 0.6 NA (10\% solid angle) lens and imaged onto an EMCCD camera.

A portion of the \SI{650}{\nano\meter} laser, controlled by an AOM, was fiber coupled into the photonic chip and fixed in place with epoxy.
We measure a combined waveguide output power of \SI{3}{\micro\watt} which is directed to the ions using the scheme shown in Fig. \ref{fig:iondiagram}.
A net M=\magnification{0.1} lens system projects an image of the outputs onto the trapped ions at a spacing matching the ions.
The dove prism allows for the photonic chip image to be rotated to match the tilt of the trapped ion chain.
The image captured in the inset of Fig. \ref{fig:iondiagram} shows a simultaneous display of the ions and the purposefully offset chip light, illustrating the ability to match ion and waveguide spacing.

\subsection{\label{sec:ion:result}Ion Demonstration Results}
The ions fluoresced and remained crystallized as the chip illuminated the ions with \SI{650}{\nano\meter} imaged light.
When the waveguide outputs translate along the ion chain, the ions encounter the dark regions between the outputs, the repumping process stops, and the ions no longer fluoresce.
If left in this state where laser cooling is unable to continue, the chain eventually decrystallizes.
Because the outputs are unevenly spaced, away from the design position, not all ions will be illuminated at once.
Quickly and continuously translating the outputs across thus causes individual ions to flicker out of sync with each other as they encounter different regions of the chip output.
This demonstrates that the individual output are well-focused in the plane of the ions.

Precise free-space optics are difficult to use with this system as the sharp focus needed to achieve good crosstalk makes alignment challenging.
Furthermore, direct fluorescence imaging limits the signal-to-noise ratio \cite{noek:2013} to much less than the light extinction measured in this work.
Thus, a measurement of the waveguide output profile using the ions as a probe is limited.
However, it is useful to demonstrate one envisioned use of the photonic circuit by directly addressing the \repumper{} repumping transition of Ba$^+$.
The ions are illuminated by the individual beams from the chip, providing the only repumping light during imaging.
The resulting fluorescence is captured in the image on the right side of Fig. \ref{fig:PrimaryFigure}.
Again, this is not intended as a quantitative measurement of the output profile.
Nonetheless, it demonstrates the viability of an intended application of the chip as a way to direct laser light to irregularly spaced targets since this method maintained a chain of cooled ions.

\section{\label{sec:conclusion}Conclusions}
We have developed a CMOS-foundry-compatible low-crosstalk photonic waveguide array, where the fabrication process allows a precise, irregular positioning of the laser outputs.
The design implemented stray light reduction techniques including trenching between outputs and turning to shed weakly-guided slab modes.
These improvements are evident in the photonic chip's measured crosstalk of \crosstalk{} using a scanning slit to evaluate low intensity signals.
This is an order of magnitude greater than the otherwise lowest crosstalk reported in literature.
A complementary 2D directly coupled fiber scan showed a \fibercrosstalk{} difference between the peak height and noise floor.
A demonstration of one application of this device allowed for direct cooling and crystallzation of a chain of trapped Ba\textsuperscript{+} ions in a single trapping region.

This approach enables the creation of modular system architectures due to its reproducible, foundry-compatible production.
By providing interfaces between qubits and optical fibers, they enable quantum networking capabilities.
The precision, accuracy, and customizability afforded by this approach can be tailored to the target application.
While the devices in this work are passive, the process supports the implementation of active devices, which can aid in directly addressing independent targets, such as qubits, and offers a scalable method to create more powerful quantum systems.
Using the mode-matching considerations in this work, these devices can also provide tailored light collection.
Combined, these capabilities can bring photon-based quantum information processing techniques to any targeted qubit platform.
Overall these characteristics make this a promising platform for interfacing with qubits.

\section{\label{sec:ack}Acknowledgements}
N.J.B., A.C.K., and A.B. acquired the waveguide characterization data, C.L.C. analyzed the data, V.S.S.S. designed the waveguides, and AIM Photonics fabricated the photonics chips. All authors contributed to this work.

This material is based on research sponsored by the United State Air Force (FA8750-23-C-1001) and Air Force Research Laboratory under AIM Photonics (agreement number FA8650-21-2-1000) and also FA8750-21-2-0004. The U.S. Government is authorized to reproduce and distribute reprints for Governmental purposes notwithstanding any copyright notation thereon. The views and conclusions contained herein are those of the authors and should not be interpreted as necessarily representing the official policies or endorsements, either expressed or implied, of the United States Air Force, the Air Force Research Laboratory or the U.S. Government. Approved for Public Release; Distribution Unlimited: PA\#: AFRL-2024-3327

\providecommand{\noopsort}[1]{}\providecommand{\singleletter}[1]{#1}%

\end{document}